\documentclass[aps,showpacs,twocolumn]{revtex4}

\usepackage[latin1]{inputenc} 
\usepackage{graphicx}
\usepackage{amsmath}
\usepackage{color}

\newcommand{\ket}[1]{\left|{#1}\right\rangle}

\def\dblone{\hbox{$1\hskip -1.2pt\vrule depth 0pt height 1.6ex width 0.7pt\vrule depth 0pt height 0.3pt width 0.12em$}}
\begin{document}
\title[quantum information storage]{Adiabatic refocusing of nuclear spins in Tm$^{3+}$:YAG }
\author{R. Lauro, T. Chaneli\`ere and J.- L. Le Gou\"et}
\email[electronic mail: ]{jean-louis.legouet@lac.u-psud.fr}
\affiliation{Laboratoire Aim\'e Cotton, CNRS UPR3321, Univ. Paris Sud, bat\^iment 505,
campus universitaire, 91405 Orsay, France}
\keywords{quantum information, storage}
\pacs{42.50.Ex,42.50.Gy,42.50.Md,82.56.Jn,03.65.Wj}

\begin{abstract}
We show that the optical absorbance detection of nuclear spin echo gives direct access to the spin concentration, unlike most coherent signal detection techniques where the signal intensity/amplitude is difficult to connect experimentally with the spin concentration. This way we measure the spin refocusing efficiency in a crystal of Tm$^{3+}$:YAG. Given the large inhomogeneous broadening of the spin transition in this material, rephasing the spins with the usual hard pulse procedure would require excessively high radiofrequency power. Instead we resort to an adiabatic pulse sequence that perfectly returns the spins to their initial common orientation, at low power cost.   
\end{abstract}
\volumeyear{year}
\volumenumber{number}
\issuenumber{number}
\eid{identifier}
\date[Date text]{date}
\received[Received text]{date}

\revised[Revised text]{date}

\accepted[Accepted text]{date}

\published[Published text]{date}

\startpage{1}
\endpage{ }
\maketitle
\section{Introduction}
The optical detection of magnetic resonance has been investigated for a long time, starting with atomic vapors~\cite{kastler1950,kastler1951,brossel1952}. In solids, the earliest studies were reported in ruby~\cite{geschwind1959}. Since the emergence of tunable lasers, special attention has been paid to the Raman heterodyne optical detection of coherent transient spin processes~\cite{mlynek1983}. Detection of coherent magnetic resonance features through optical fluorescence was examined even earlier~\cite{breiland1973}. In the latter class of processes, a $\pi$/2 RF pulse is used to convert the spin coherence into a level population that can emit light by fuorescence. A similar approach was also used to detect spin echoes through optical absorbance~\cite{shelby1978}. 

Optically detected magnetic resonance is enjoying renewed interest as research on quantum memory for light becomes more popular. Most quantum storage schemes rely on the conversion of optical atomic coherence into long lifetime spin coherence and back~\cite{fleischhauer2000,moiseev2001,nilsson2005,lauro2009a,afzelius2009}. Echo techniques are needed to compensate for the spin inhomogeneous phase shift~\cite{turukhin2001,longdell2005}, i.e. to refocus the spins. 

In the present paper our aim is twofold. First we examine the potential of adiabatic rapid passage (ARP) for spin refocusing, when hard pulses would require excessive RF power. Second, placing this research in the context of quantum memory, we explore the specificity of magnetic resonance detection through the absorption of light. We show that this approach may convey more information than the direct detection of coherent emission. This relies on the fact that an absorption coefficient is related to the concentration of absorbing centers in a simple and precise way. 

After discussing the principle of adiabatic spin refocusing in Section II, we present the absorbance detection of spin refocusing in Section III and IV. The experiment, described in Section V, is conducted in Tm$^{3+}$:YAG, an actively investigated system in the prospect of quantum storage~\cite{deSeze2006,louchet2007,louchet2008,lauro2009,lauro2009a,chaneliere2010,bonarota2010}. 
          
\section{Adiabatic spin refocusing}\label{adiabatic refocusing}
When spins interact with a spatially non-uniform static vertical magnetic field, the energy level spacing is inhomogeneously broadened. Then, excitation by a RF horizontal pulse gives rise to an oscillating magnetization that rapidly vanishes as the various spins precess at different rates around the static field. It is well known that a $\pi$-pulse can rephase the spins and restore the oscillating horizontal magnetization. In order to uniformly excite all the spins, this pulse must be shorter than the inverse inhomogeneous broadening. This condition may entail inaccessible or unrealistic RF power requirements. Indeed, at given pulse area, the needed RF power varies as the square of the inverse pulse duration. By successively driving the spins with different energy level spacings, frequency-chirped RF pulses may offer a less power-demanding way of excitation. This is precisely the way ARP works.

Since the early days of Nuclear Magnetic Resonance (NMR), ARP has been extensively used to flip vertically oriented spins, or equivalently, to perform a total transfer of population between two energy levels~\cite{abragam1961}. This works as well on optical transitions~\cite{loy1974}. Adiabatic passage is generally performed by sweeping the RF field frequency, while the static field remains constant. The spin is driven by an effective magnetic field $\mathbf{B}_{eff}$ whose horizontal and vertical components are repectively equal to the field amplitude and proportional to the detuning. As the field frequency is swept, $\mathbf{B}_{eff}$ rotates, flipping for instance from downward to upward orientation if the detuning is varied from negative to positive values. The spin adiabatically follows $\mathbf{B}_{eff}$ provided the rotation speed is much smaller than the precession rate of the spin around $\mathbf{B}_{eff}$. Vertical flip or population transfer, the most popular ARP application, does not preserve any phase-shift since the initial state is orthogonal to the final one. On the contrary, spin refocusing aims at making the final superposition state coincide with the initial one. The phase control is essential. A single ARP is not enough but a pair of ARP pulses can refocus the spins in any direction~\cite{conolly1991,degraaf1997,garwood2001}. 

\begin{figure}
  \centering
  {\includegraphics[width=\columnwidth]
{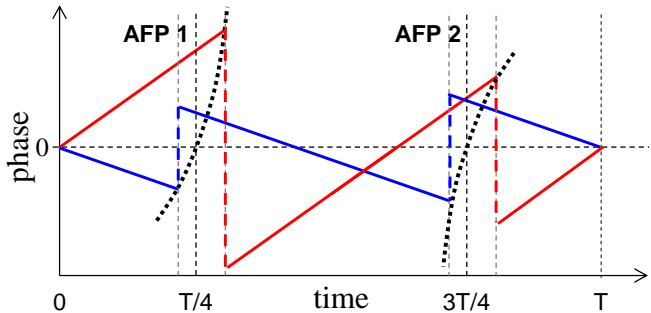}}
  \caption{(color online) Time evolution of spin angular positions in the horizontal plane. The red and blue lines correspond to two spins with different detunings. Initially aligned, the two spins return to their initially orientation at time $T$, after undergoing two ARP excitations centered at times $T/4$ and $3T/4$. The black dotted lines represent the locus of the flipping positions.}
  \label{fig:refocusing_principle}
\end{figure}

The principle of spin refocusing by a pair of ARP pulses is illustrated in Fig.~\ref{fig:refocusing_principle}. The figure represents the evolution of the phase, i.e. the angular position, of two spins in the horizontal plane. The two spins differ by their detuning with respect to the center $\omega_0$ of the inhomogeneously broadened energy spacing distribution. They are aligned at initial time $t=0$. Then rotating at different rates around the vertical axis, they depart from each other. One applies a first ARP pulse whose frequency is swept at rate $r$ through $\omega_0$  at time $T/4$. The pulse gets resonant with a spin of frequency $\omega$ at time $T/4+(\omega-\omega_0)/r$, when the phase equals $rt(t-T/4)$. At that moment the spin vertical component is flipped and the horizontal angular position is reverted. In the figure, flipping is assumed to last much less than $T/4$. Then the spin phase continues to evolve at the same rate until interaction with a second ARP pulse, identical to the first one, and centered at time $3T/4$. The interaction time is shifted by the same amount $(\omega-\omega_0)/r$ with respect to $3T/4$. This second pulse reverts again the phase and restores the initial vertical spin component. Finally the spin are phased together again at time $T$. The locus of the flipping positions in the phase/time representation is comprised of two quadratic segments defined by $\phi=rt(t-T/4)$ and $\phi=r(T-t)(t-3T/4)$.       

\begin{figure}
  \centering
  {\includegraphics[width=\columnwidth]
{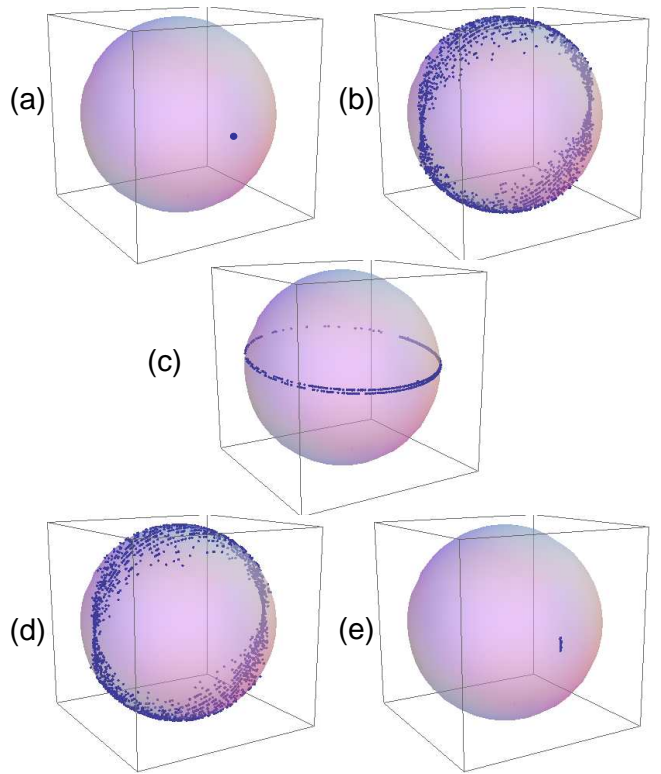}}
  \caption{(color online) Simulation of spin refocusing. The spins, distributed over a 1MHz-wide inhomogeneous distribution, are all directed along $Ox$ at time $t=0$. Their position on the unit sphere is represented at times 0 (a), $T/4$ (b), $T/2$ (c), $3T/4$ (d) and $T$ (e), where $T=0.2$ms. Two ARPs are applied successively, for 0.1ms each. They are frequency-chirped at the constant 0.04MHz/$\mu$-rate. They are swept through the center of the inhomogeneous distribution at times $T/4$ and $3T/4$ respectively. Their Rabi frequency is kept fixed at 0.28MHz.}
  \label{fig:spheres}
\end{figure}

A simulation of spin refocusing by a pair of successive ARPs is presented in Fig.~\ref{fig:spheres}. Each ARP lasts for $T/2=$0.1ms, during which the RF field Rabi frequency is maintained fixed at 0.28MHz. The RF is swept at a constant rate of 0.04MHz/$\mu$s. The 2000 spins are distributed over a 1 MHz-wide gaussian inhomogeneous distribution. At time $t=0$, they are all directed along axis $Ox$. The figure illustrates the strong scattering of spin directions at times $T/4$, $T/2$ and $3T/4$, strikingly contrasting with refocusing at time $T$.

\section{Measuring the spin refocusing efficiency}\label{refocusing efficiency}
Aiming at measuring the spin refocusing efficiency, we have to compare the final oriented spin concentration with the initial one. This information cannot be provided by a standard spin echo experiment. Indeed, it is difficult to relate the echo intensity to the initial spin coherence concentration. Usually an echo experiment only gives access to the signal intensity dependence on various parameters such as the pulse delay, temperature, external fields etc... Optical absorption is a simple tool to reach the desired quantity. 

\begin{figure}[h!]
\centering
\includegraphics[width=4cm]{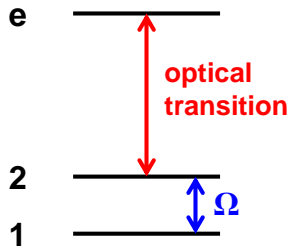}
\caption{Optical probing of magnetic resonance. The upper electronic state $\ket{e}$ is optically connected to $\ket{2}$, a nuclear spin sublevel of the electronic ground state. As the $\ket{1}-\ket{2}$ spin transition is excited by the RF field at Rabi frequency $\Omega$, the state $\ket{2}$ population is optically probed by absorption on the $\ket{2}$-$\ket{e}$ transition.}
\label{fig:measurement}
\end{figure}

Let the two states $\ket 1$ and $\ket 2$ of a $I=1/2$ nuclear spin be connected by optical transitions to a common $\ket e$ excited state (see Fig.\ref{fig:measurement}). The spins are first prepared in state $\ket 1$ by optical pumping from state $\ket 2$. In other words, all the spins are aligned vertically, with the same orientation. The transmitted optical intensity is originally measured as $I_1$ on transition $\ket 2 \leftarrow \ket e$. As the spins are pumped from $\ket 2$ to $\ket 1$, the sample absorption vanishes and the transmitted intensity raises to the input value $I_0$.   

At time $t=0$, one applies a RF $\pi/2$ hard pulse to rotate the spins into the horizontal plane. Following the pulse, the vertical spin component drops to 0, as can be monitored through the absorption increase on transition $\ket 2 \leftarrow \ket e$. States $\ket 1$ and $\ket 2$ are equally populated, which actually makes the absorption coefficient recover its original thermal equilibrium value. The transmitted intensity drops to $I_1$.  

A $\pi/2$ hard pulse, applied immediately after the first one but with opposite sign, should be able to rotate back the spins to their original orientation. Instead, one lets the spins precess freely after the first $\pi/2$ pulse. Initially aligned, the spins begin to depart from each other in the horizontal plane, until they are evenly oriented in all directions. This occurs in a time interval equal to the inverse inhomogeneous width. After the spin direction is randomized, excitation by a $\pi/2$ hard pulse is no longer able to rotate back the spins to state $\ket 1$ and to make the absorption vanish on transition $\ket 2 \leftarrow \ket e$. The transmitted intensity remains equal to $I_1$.

To make the absorption sensitive again to a $\pi/2$ pulse, one refocuses the spins with a pair of ARP full pulses that shine the sample at times $T/4$ and $3T/4$. At time $T$, the spins are aligned again. A $\pi/2$ pulse is used to rotate back the spins to vertical position. However, the horizontal spin component may have relaxed during the time interval $T$. If the spins do not totally return to state $\ket 1$, absorption does not vanish. With exponential relaxation, at rate $\gamma$, the final absorption on transition $\ket 2 \leftarrow \ket e$ is reduced by factor $1-\mathrm{e}^{-\gamma T}$ with respect to the original value, in the medium at thermal equilibrium.

\section{Replacing $\pi$/2-hard pulses by adiabatic half passage}\label{adiabatic half passage} 
Resorting to $\pi/2$ hard pulses is somehow inconsistent with previous assumptions. As already discussed, the uniform excitation of the inhomogeneously broadened spin transition would demand very brief hard pulses with unacceptably high power. Adiabatic passage may offer an alternative way to rotate the spin into the horizontal plane. Let us consider the transformation operated by an adiabatic half passage (AHP) process ~\cite{degraaf1997}, an ARP that is interrupted at time $t=0$, when the pulse frequency is scanned through $\omega_0$. We restrict the discussion to the most simple case where the RF field amplitude is kept constant over the pulse duration, although more sophisticated adiabatic schemes, such as the complex hyperbolic secant, turn out to be more popular in NMR and optical~\cite{deSeze2005,rippe2005} applications. 

According to Appendix~\ref{ss:AHP_effect}, the AHP turns the initially vertical spin $M(\Delta,-\infty)=(0,0,1)$ into:
\begin{equation}
M(\Delta,0)=\frac{1}{\sqrt{\Omega^2+\Delta^2}}(-\Omega,0,\Delta)
\end{equation} 
where the Rabi frequency $\Omega=\gamma_BB$ is expressed in terms of the RF field constant amplitude $B$ and the gyromagnetic factor $\gamma_B$. The spins with large negative $\Delta$ detuning undergo an upside down flip, while those with large positive detuning preserve their initial orientation. Hence $M_z(\Delta,0)$ is an \textit{odd} function of $\Delta$. Spins with $|\Delta|<\Omega$ are rotated into the horizontal plane, irrespective of the detuning sign. 

Let the spins be distributed over a $\Delta_0$-broad inhomogeneous width, according to an \textit{even} distribution $G(\Delta)$, with unit normalization $\int G(\Delta)\mathrm{d}\Delta=1$. Then the absorption coefficient on transition $\ket 2 \leftarrow \ket e$ reads as
\begin{equation}
\alpha(t)=\alpha_0\int G(\Delta)\left[1-M_z(\Delta,t)\right]\mathrm{d}\Delta
\label{eq:def_alpha}
\end{equation} 
At equilibrium, $M_z(\Delta,-\infty)=0$, and $\alpha(-\infty)=\alpha_0$. After pumping from $\ket 2$ to $\ket 1$, $\alpha(t)$ vanishes. The AHP makes $\alpha(t)$ change back to $\alpha_0$ at $t=0$, exactly as a $\pi/2$ hard pulse would do, the transmitted intensity returning to $I_1$.  

However, the vertical spin component only vanishes on average. When the refocusing AFP pulses excite the medium, the spins with different $\Delta$ values do not flip simultaneously, which is reflected by a transient variation of absorption. Let the spins evolve freely until the first AFP pulse, centered at time $T/4$, with $\Delta_0 T/4>>1$. As far as absorption is concerned, inhomogeneous dephasing enables us to ignore the conversion of the horizontal components into the vertical one. Indeed this contribution vanishes when averaged over $\Delta$. According to Appendix~\ref{ss:AFP_effect}, $M_z(\Delta,t)$ varies as:
\begin{equation}
M_z(\Delta,t)=\frac{\Delta}{\sqrt{\Omega^2+\Delta^2}}\frac{\Delta -\dot{\phi}(t-T/4)}{\sqrt{\Omega^2+\left[\Delta -\dot{\phi}(t-T/4)\right]^2}}   
\label{bump_profile}      
\end{equation}
The resulting variation of $\alpha(t)$, given by Eq.~(\ref{eq:def_alpha}), carries information on $G(\Delta)$. After reaching a minimum: 
\begin{equation}
\alpha(T/4)=\alpha_0\int \frac{G(\Delta)\Omega^2}{\Omega^2+\Delta^2}\mathrm{d}\Delta
\label{eq:alpha_t2}
\end{equation}  
at $T/4$, $\alpha(t)$ returns to $\alpha_0$, while the transmitted intensity reaches a maximum $I_2$ before dropping back to $I_1$. In a similar way, the transmitted intensity decreases during the second AFP pulse at $3T/4$.  

At time $T$ the spins are refocused as:
\begin{equation}
M(\Delta,T)=\frac{1}{\sqrt{\Omega^2+\Delta^2}}(-\Omega\mathrm{e}^{-\gamma T},0,\Delta)
\label{eq:spinAHP}
\end{equation}
where the transverse relaxation has been accounted for. An AHP is used again to rotate the spins back to vertical orientation. The pulse frequency is swept with opposite rate with respect to the first AHP, starting at frequency $\omega_0$ at time $T$. The final $M_z$ component reads as:
\begin{equation}
M(\Delta,+\infty)=\frac{\Omega^2\mathrm{e}^{-\gamma T}+\Delta^2}{\Omega^2+\Delta^2},
\end{equation}
which leads to the final absorption:
\begin{equation}
\alpha(T)=\alpha_0\left(1-\mathrm{e}^{-\gamma T}\right)\int \frac{G(\Delta)\Omega^2}{\Omega^2+\Delta^2}\mathrm{d}\Delta
\label{eq:alpha_final}
\end{equation}
The final intensity $I_f(T)$ is related to $I_0$ and $I_2$ by:
\begin{equation}
\ln(I_f(T)/I_2)=\mathrm{e}^{-\gamma T}\ln(I_0/I_2)
\label{eq:I_final}
\end{equation}
Should horizontal refocusing be imperfect, $\mathrm{e}^{-\gamma T}$ should be replaced by $\eta\mathrm{e}^{-\gamma T}$ where $\eta<1$. Hence, the refocusing efficiency can be defined as:
\begin{equation}
\eta=\mathrm{e}^{\gamma T}\ln(I_f(T)/I_2)/\ln(I_0/I_2)
\label{eq:efficiency}
\end{equation}
Provided $\eta$ does not vary with $T$, Eqs.~(\ref{eq:I_final}) and (\ref{eq:efficiency}) can be recombined into:
\begin{equation}
\ln\left(I_f(T)/I_f(\infty)\right)=\mathrm{e}^{-\gamma(T-T_0)}\ln\left(I_f(T_0)/I_f(\infty)\right),
\label{eq:mesure_gamma}
\end{equation}
Hence, the variations of $\ln\left(I_f(T)\right)$ with $T$ give access to the $\gamma$ decay rate, irrespective of $\eta$ value.

\section{Experimental}

The experiments are carried out in a 0,1 \% at. Tm$^{3+}$:YAG crystal. An external magnetic field lifts the nuclear spin degeneracy. The resulting 4-level structure is comprised of two ground states and two excited states. We choose the same field orientation as in Ref.~\cite{louchet2007}.

The optical aspects of the setup have been described extensively in Refs.\cite{louchet2007,lauro2009}. Basically, the light beam, emerging from an extended cavity diode laser, is amplitude and phase-shaped by acousto-optics modulators, driven by a high sample-rate arbitrary waveform generator (AWG). The crystal is cooled-down to 1,7 K in a liquid Helium cryostat. The static magnetic field is generated by superconductive coils and is set to about 0,5 T, which leads to a ground state splitting of 15,7 MHz. The spin transition is resonantly driven by a RF magnetic field. This excitation is conveyed to the crystal by a 10-turn, 20 mm long, 10mm in diameter, coil oriented along the light pulse wave vector. The crystal sits at the coil center. The RF signal, generated by the AWG, is fed to the coil through  a 500 W amplifier.

First we calibrate the Rabi frequency of the spin transition, as a function of the voltage applied to the coil. To this end, we optically detect the Rabi oscillations of the spin states. Let's go back to the system described in Fig.\ref{fig:measurement}. After optically pumping the ions in state $\ket{1}$, we apply the RF excitation and detect the state oscillations by monitoring the optical absorption on transition $\ket{2}$-$\ket{e}$ with the help of a weak probe beam. Provided the probe intensity is low enough, the absorption coefficient is proportional to the population of $\ket{2}$ and oscillates at the Rabi frequency, 
\begin{figure}[h!]
\centering
\includegraphics[width=\columnwidth]{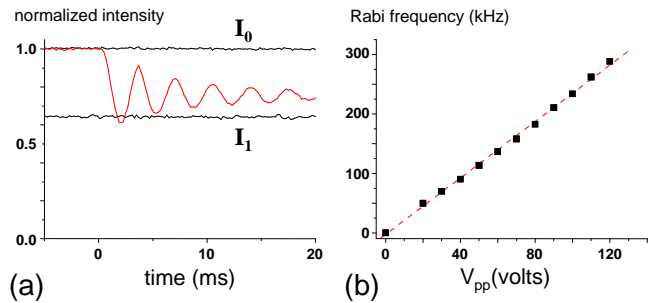}
\caption{(color online) Nutation experiment. (a) time evolution of the normalized transmission through the crystal. The transmitted optical intensity, equal to I$_1$ when both ground state sublevels are equally populated, raises to I$_0$ when all spins are prepared in state $\ket{1}$. When the RF magnetic-field is applied, the transmitted intensity oscillates (red line). The RF field is turned on at t=0. The peak-to-peak voltage at the coil is set to 120 V. (b) Rabi frequency as a function of the applied peak-to-peak voltage.}\label{fig:nutation}
\end{figure}
as observed in Fig.~\ref{fig:nutation}(a). They are damped because the spins are not uniformly coupled to the RF field. This partly results from the inhomogeneous broadening of the spin transition, and from the $\Delta$ detuning dependence of the $\sqrt{\Omega^2 + \Delta^2}$ effective Rabi frequency. This also reflects the non-uniformity of the RF field over the crystal, given the finite length of the coil and the field distorsions by metallic parts of the cryostat. The excitation non-uniformity also explains the limited amplitude of the oscillations in Fig.~\ref{fig:nutation}(a). Total inversion is not observed. From the data in Fig.~\ref{fig:nutation}(b), $\Omega$ is measured to vary linearly with the peak-to-peak voltage, measured at the coil input, at the rate of 2.4$\pm$0.1 kHz/V.    

Next we turn to the adiabatic spin refocusing experiment. We use the most simple adiabatic passage procedure where a fixed-amplitude field is frequency-chirped at constant rate $r$. As pointed out in Section~\ref{adiabatic refocusing}, the precession speed of the spin around the driving vector, $\Omega$, shall be much larger than the rotation speed of the driving vector, $r/\Omega$. In addition, the duration of the pulse shall by far exceed the time $\Delta_0/r$ needed to scan the inhomogeneous width $\Delta_0$. Finally, since an ARP is a coherent process, the flipping time $\Omega/r$ shall be much smaller than the spin coherence lifetime. The adiabatic pulse (AFP) features are summarized in Tab. \ref{table:parameters}. 
\begin{table}[h]
   \begin{center}
\begin{tabular}{|c|c|c|c|c|}
  \hline
  $\Omega$/(2 $\pi$) & $r$/(2 $\pi$) & duration & $\Omega^2/r$ & $\Omega/r$ \\
  \hline
  284,4 kHz & 40 kHz/$\mu$s & 100 $\mu$s & 12,7 & 7.1 $\mu$s \\
  \hline
\end{tabular}
 \end{center}
  \caption{Features of the adiabatic pulses used for spin refocusing}
  \label{table:parameters}
\end{table} 
\begin{figure}
\includegraphics[width=7.1cm]{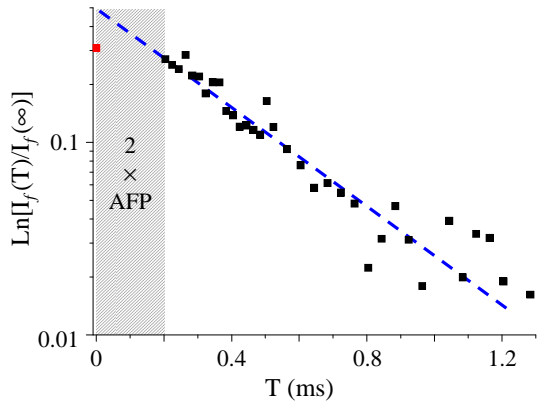}
\caption{(color online) Measurement of $\ln\left(I_f(T)/I_f(\infty)\right)$ as a function of delay $T$. Exponential decay fit leads to $\gamma^{-1}=0.33\pm0.05$ms. The dashed (red) line corresponds to exponential decay at rate $\gamma=3.0$ms$^{-1}$. The shaded area is inaccessible. Indeed, the pair of refocusing AFPs prevents us from reducing $T$ under 200$\mu$s. However, refocusing is not needed if one applies the second AHP just after the first one. From the measurement of $\ln\left(I_f(T)/I_f(\infty)\right)$ at $T=0$ (red dot), one infers that $I_f(T)$ decays at rate smaller than $\gamma=3.0$ms$^{-1}$ from $T=0$ to $T=0.2$ms.}
\label{fig:transverse_decay}
\end{figure}
They are expected to satisfy the three adiabatic passage conditions. The spin coherence lifetime can be deduced from the $T$-dependence of the probe beam transmitted intensity $I_f(T)$, measured at the end of the second AHP (see Eq.~(\ref{eq:mesure_gamma})). According to the experimental data, displayed in Fig.~\ref{fig:transverse_decay}, $\ln\left(I_f(T)/I_f(\infty)\right)$ decays exponentially as expected, which leads to a spin coherence lifetime of $\gamma^{-1}=0.33\pm0.05$ ms. This result is consistent with a previous Raman echo measurement~\cite{louchet2008}. The exponential decay is also consistent with photon echo experiments that were performed in a Tm$^{3+}$:YAG crystal at the same concentration, and in the same temperature conditions where the relaxation of both optical and spin coherences is dominated by spin-spin interaction~\cite{macfarlane1993}. However the spin coherence appears to decay at slower rate between $T=0$ and $T=0.2$ ms. This effect deserves further investigation.  

\begin{figure}
\includegraphics[width=7.1cm]{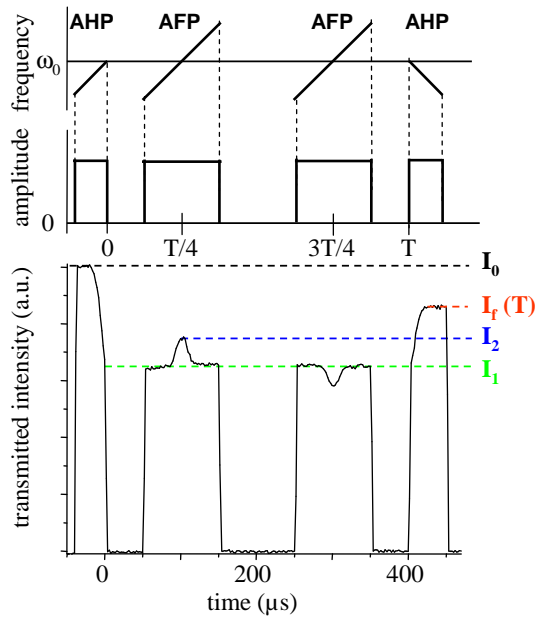}
\caption{(color online) spin refocusing experiment in a 0.1 at. \% Tm$^{3+}$:YAG crystal. Upper box: amplitude and frequency variations of the RF driving field. Lower box: transmitted optical intensity. After optical transfer to state $\ket 2$ the medium is transparent ($I_0$). The first AHP rotates the spins and restore the equilibrium transmission ($I_1$). The transmitted intensity raises to a maximum $I_2$ during the AFP and behaves in the opposite way during the second AFP. At the end of the second AHP the transmitted intensity reaches its final $I_f(T)$ value.}
\label{fig:refocusing_experiment}
\end{figure}

The refocusing efficiency can now be measured. The time-evolution of the probe beam transmitted intensity is shown in Fig.~\ref{fig:refocusing_experiment}, together with the RF adiabatic sequence.         

\begin{figure}
\includegraphics[width=7.1cm]{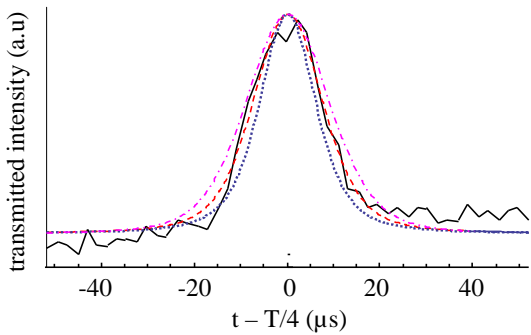}
\caption{(color online) transmission bump at T/4. The transmitted optical intensity temporal profile, as derived from Eq.~(\ref{bump_profile}), is fitted to the experimental data, assuming a gaussian inhomogeneous distribution. The RF frequency is swept at a rate of 40MHz/ms, and the Rabi frequency has been measured to be $\Omega/(2\pi)=$0.28MHz. The dotted (blue), dashed(red) and dotted-dashed(magenta) lines respectively correspond to an inhomogheneous bandwidth ( full width at half maximum) of 0.3, 0.5 and 0.7 MHz.}
\label{fig:measuring_inhom_width}
\end{figure}

In accordance with the storyline of Section~\ref{adiabatic half passage}, the incoming probe intensity $I_0$ is totally transmitted at the beginning, through an intially transparent material. The first AHP excitation rotates the spins into the horizontal plane, which makes the transmitted intensity drop to $I_1$ at time 0. Then, during the AFP refocusing steps, the transmitted intensity remains close to $I_1$, except around times $T/4$ and $3T/4$. It first exhibits a bump, reaching the maximum value $I_2$ at times $T/4$, then an anti-bump at time $3T/4$. Those variations correspond to the flip of the vertical spin component. Indeed only the spins at $\omega_0$ have been perfectly rotated into the horizontal plane by the AHP. Off-resonant spins, distributed over the $\ket 1$-$\ket 2$ transition inhomogeneous width $\Delta_0$, are only partially rotated and keep a vertical component. A magnified view of the bump at $T/4$ is shown in Fig.~\ref{fig:measuring_inhom_width}. We determine $\Delta_0$ (full width at half maximum) by fitting the bump profile with Eq.~(\ref{bump_profile}). We obtain $\Delta_0=0.5\pm0.1$ MHz, which is consistent with previous measurements~\cite{louchet2008}. 

After interaction with the two AFPs, the spins are expected to be aligned together at time $T$. At that moment, after changing the chirp rate sign, we use a second AHP to rotate back the spins to their initial vertical orientation. Perfect refocusing, with no decoherence, should make the transmitted intensity return to $I_0$. The measured transmitted intensity is denoted $I_f(T)$. Substituting the Fig.~\ref{fig:refocusing_experiment} experimental data and $\gamma^{-1}=0.33 ms$ into Eq.~(\ref{eq:efficiency}), we obtain $\eta=1.6\pm 0.3$. This unphysical value, much larger than the unit upper boundary, suggests that the spin coherence relaxation might be slower than expected. We suspect that the 0.2 ms interval between the adiabatic passages is not much larger than the reservoir fluctuation correlation time. As a consequence, the spin decoherence might be partially suppressed by the two successive AFPs~\cite{herzog1956,mims1968}. This is confirmed by experimental data in Fig.~\ref{fig:transverse_decay}.  

\section{Conclusion} 
We have used optical detection to investigate the spin refocusing capabilities of adiabatic pulses in a Tm$^{3+}$:YAG crystal. As in any conventional spin echo experiment we have measured the transverse relaxation time of the Tm$^{3+}$ ion spin coherence in the electronic ground state. Moreover we have shown that the optical absorbance procedure gives direct access to the spin coherence concentration. As a consequence, one can measure the spin refocusing efficiency. Spin decoherence seems to be slowed down on time scales shorter than, or on the order of, 0.2 ms. If confirmed, this system might offer an interesting test bed for extending the Carr-Purcell Meiboom-Gill~\cite{meiboom1958,longdell2005} method to adiabatic passages. 

This work is supported by the European Commission through the FP7 QuRep project, and by the national grant ANR-09-BLAN-0333-03

\appendix
\renewcommand\theequation{A.\arabic{equation}} 
\section{Adiabatic Rapid Passage description}\label{ARP_theory}
\subsection{ARP operator}
This very standard calculation is presented for the sake of completeness. In addition we want to point out the importance of a common reference frame in a problem where the medium undergoes successive excitations by RF fields $B_i(t)$. Those fields are swept through $\omega_0$ at different times $t_i$. They can be expressed as:
\begin{equation}
B_i(t)=B\cos\left[\omega_0t+\phi(t-t_i)\right]
\label{eq:ARP_field_def}
\end{equation}
Instead of changing immediately to the frame rotating at the instantaneous frequency of the field, we first make a change to the common reference frame rotating at frequency $\omega_0$. In this frame, the rotating wave approximation of the Bloch equation reads as:
\begin{equation}
\left\{\begin{array}{ll}
\partial_tM_x=-\Delta M_y-\Omega\sin\left[\phi(t-t_i)\right]M_z\\                   
\partial_tM_y=\Delta M_x+\Omega\cos\left[\phi(t-t_i)\right]M_z\\
\partial_tM_z=\Omega\sin\left[\phi(t-t_i)\right]M_x-\Omega\cos\left[\phi(t-t_i)\right]M_y
\end{array}\right. 
\end{equation}
where $\Delta=\omega_{12}-\omega_0$ and the Rabi frequency $\Omega=\gamma_BB$ is expressed in terms of the RF field constant amplitude $B$ and the gyromagnetic factor $\gamma_B$. One performs two successive changes of frame. First one transforms the spin coordinates according to the rotation $M=R_1^{(i)}M'$ where:
\begin{align}
R_1^{(i)}(t)= \begin{pmatrix} \cos\left[\phi(t-t_i)\right]&-\sin\left[\phi(t-t_i)\right]&0\\
\sin\left[\phi(t-t_i)\right]&\cos\left[\phi(t-t_i)\right]&0\\0&0&1 \end{pmatrix}
\label{eq:Rot_1}
\end{align} 
The new frame rotates in concert with the frequency swept field, and the Bloch equations are changed into:
\begin{equation}
\left\{\begin{array}{ll}
\partial_tM_{x'}=-\left[\Delta -\dot{\phi}(t-t_i)\right]M_{y'}\\                   
\partial_tM_{y'}=\left[\Delta -\dot{\phi}(t-t_i)\right] M_{x'}+\Omega M_{z'}\\
\partial_tM_{z'}=-\Omega M_{y'}
\end{array}\right. 
\end{equation} 
To align the $z"$ axis along the driving vector, one performs a second transformation that is defined by the rotation $M'=R_2^{(i)}M''$, where:
\begin{align}
R_2^{(i)}(t)= \begin{pmatrix} \cos\Theta^{(i)}&0&-\sin\Theta^{(i)}\\
0&1&0\\
\sin\Theta^{(i)}&0&\cos\Theta^{(i)}
 \end{pmatrix}
\label{eq:Rot_2}
\end{align}  
where: 
\[
\sin\Theta^{(i)}=\Omega/\Omega_{eff},\;\cos\Theta^{(i)}=\left[\Delta -\dot{\phi}(t-t_i)\right]/\Omega_{eff},
\] and
\[
\Omega_{eff}(t)=\sqrt{\Omega^2+\left[\Delta -\dot{\phi}(t-t_i)\right]^2}.
\]
In this frame the Bloch equation reads as:
\begin{equation}
\left\{\begin{array}{ll}
\partial_tM_{x''}-M_{z''}\dot{\Theta}^{(i)}=-\Omega_{eff}M_{y''}\\                   
\partial_tM_{y''}=\Omega_{eff}M_{x''}\\
\partial_tM_{z''}+M_{x''}\dot{\Theta}^{(i)}=0
\end{array}\right. 
\end{equation}
Assuming adiabatic conditions are satisfied, one can neglect $M_{z''}\dot{\Theta}^{(i)}$ and $M_{x''}\dot{\Theta}^{(i)}$, which leads to:
\begin{equation}
\left\{\begin{array}{ll}
\partial_tM_{x''}=-\Omega_{eff}M_{y''}\\                   
\partial_tM_{y''}=\Omega_{eff}M_{x''}\\
\partial_tM_{z''}=0
\end{array}\right. 
\end{equation}
The solution of this equation can be expressed as the following propagator:
\begin{align}
U^{(i)}(t\leftarrow t')= \begin{pmatrix} \cos\Phi^{(i)}&-\sin\Phi^{(i)}&0\\
-\sin\Phi^{(i)}&\cos\Phi^{(i)}&0\\
0&0&1
 \end{pmatrix}
\label{eq:2ndframe_Prop}
\end{align}  
where: $\Phi^{(i)}=\int_{t'}^t\Omega_{eff}(t'')\mathrm{d}t''$. Finally, in the common frame rotating at $\omega_0$, the effect of an ARP on the spin coordinates can be described by the propagator:
\begin{align}
V^{(i)}&(t\leftarrow t')=\notag\\
&R_1^{(i)}(t)R_2^{(i)}(t)U^{(i)}(t\leftarrow t')\left[R_1^{(i)}(t')R_2^{(i)}(t')\right]^{-1}
\label{eq:ARP_Prop}
\end{align}
\subsection{AHP effect}\label{ss:AHP_effect}
The effect of the first AHP is described by $V^{(0)}(t_0\leftarrow -\infty)$. One easily verifies that $R_2^{(0)}(-\infty)=\dblone$. As a consequence, the operator
\[
U^{(0)}(t_0\leftarrow -\infty)\left[R_1^{(0)}(-\infty)R_2^{(0)}(-\infty)\right]^{-1}
\]
has no effect on the initially vertical spin $M(\Delta,-\infty)=(0,0,1)$. One can also check that $R_1^{(0)}(t_0)=\dblone$. Hence 
\begin{equation}
M(\Delta,t_0)=R_2^{(0)}(t_0)M(\Delta,-\infty)   
\label{eq:M_t_0}
\end{equation}
where, according to Eq.~(\ref{eq:Rot_2}),
\begin{align}
R_2^{(0)}(t_0)=\frac{1}{\Omega_{eff}} \begin{pmatrix} \Delta&0&-\Omega\\
0&1&0\\
\Omega&0&\Delta
 \end{pmatrix}
\label{eq:Rot_t_0} 
\end{align} 
and $\Omega_{eff}=\sqrt{\Omega^2+\Delta^2}$. The action of the second AHP can be analyzed in a similar way.
\subsection{AFP effect on the vertical spin component}\label{ss:AFP_effect}
We consider the vertical spin component evolution in the vicinity of time $t_1$, when the first AFP frequency comes into coincidence with $\omega_0$. Since the spins have precessed for a time much longer than the inverse inhomogeneous width, their $\Delta$-averaged horizontal component vanishes. Therefore, the vertical component evolves alone and, quite in the same way as in Appendix~\ref{ss:AHP_effect}, the AFP effect is completely described by $R_2^{(1)}(t)$. More specifically, 
\begin{equation}
M_z(\Delta,t)=M_z(\Delta,t_0)\frac{\Delta -\dot{\phi}(t-t_1)}{\sqrt{\Omega^2+\left[\Delta -\dot{\phi}(t-t_1)\right]^2}}         
\end{equation}
where $M_z(\Delta,t_0)$ is given by Eq.~\ref{eq:M_t_0}.    
\section{Spin refocusing}
The RF fields $B_1(t)$ and $B_3(t)$, defined according to Eq.~(\ref{eq:ARP_field_def}), are centered at times $t_1=t_0+T/4$ and $t_3=t_0+3T/4$. Interacting with those fields, a spin precessing in a static field undergoes two successive identical ARPs. The $\Delta$ detuning is $<<\omega_0$. According to Eq.~(\ref{eq:ARP_Prop}), propagation from $t_0$ to $t_0+T$ can be expressed as $\displaystyle V^{(3)}(t_4\leftarrow t_2)V^{(1)}(t_2\leftarrow t_0)$, where $t_2=t_0+T/2$. Since the two ARPs are identical, $\displaystyle U^{(3)}(t_4\leftarrow t_2)=U^{(1)}(t_2\leftarrow t_0)$, $\displaystyle R_j^{(1)}(t_0)=R_j^{(3)}(t_2)$, and $\displaystyle R_j^{(1)}(t_2)=R_j^{(3)}(t_4)$, according to the above definitions (see Eqs~(\ref{eq:Rot_1}) and~(\ref{eq:2ndframe_Prop})). With the additional assumption $\phi(t)=\phi(-t)$, all the relevant $\displaystyle R_1^{(i)}(t_k)$ matrices are identical. It is also assumed that $\displaystyle \Omega<<|\Delta|+|\dot{\phi}(t_0-t_1)|$ so that each ARP can be completed. Then $\displaystyle R_2^{(1)}(t_0)\cong1$ and:
\begin{align}\notag
R_2^{(1)}(t_2)\cong \begin{pmatrix} -1&0&0\\0&1&0\\0&0&-1 \end{pmatrix},
\end{align}  
provided $\dot{\phi}(t_0-t_1)<0$. If $\dot{\phi}(t_0-t_1)>0$, $R_2^{(1)}(t_0)$ and $R_2^{(1)}(t_2)$ expressions should be swapped. Combining all these properties, one easily shows that 
\[
V^{(3)}(t_4\leftarrow t_2)V^{(1)}(t_2\leftarrow t_0)\cong\dblone
\]
At $t_0+T$, the three spin components return to their $t_0$ initial values.

\end{document}